\documentclass[letterpaper,10pt,aps,prd,amsmath,amssymb,floatfix,nofootinbib,reprint,superscriptaddress]{revtex4-1}

\usepackage{graphicx}  
\usepackage{multirow}
\usepackage[dvipsnames]{xcolor}
\usepackage[colorlinks=true,allcolors=BlueViolet]{hyperref}

\usepackage{latexsym}
\usepackage{amsfonts}
\usepackage{slashed}
\usepackage{color}

\usepackage{supertabular} 
\usepackage{epsfig}

\newcommand{\mbare}{\stackrel{\circ}{m}_{b\bar s}}
\newcommand{\comp}{{\rm C}\hspace{-1ex}\rule{0.1mm}{1.5ex}\hspace{1ex}}
\newcommand{\reales}{{\rm R}\hspace{-1ex}\rule{0.1mm}{1.5ex}\hspace{1ex}}
\begin{document}

\widetext



\title{\boldmath  Lowest lying  even-parity $\bar B_s$ mesons: heavy
  quark spin-flavor symmetry,  chiral dynamics,  and constituent quark model bare masses }

\author{M. Albaladejo}

\author{P. Fernandez-Soler}

\author{J. Nieves}

\author{P. G. Ortega}

\affiliation{Instituto de F\'isica Corpuscular (IFIC),
             Centro Mixto CSIC-Universidad de Valencia,
             Institutos de Investigaci\'on de Paterna,
             Aptd. 22085, E-46071 Valencia, Spain}

\date{\today}

\begin{abstract}
The discovery of the $D^\ast_{s0}(2317)$ and $D_{s1}(2460)$ resonances
in the charmed-strange meson spectra revealed that formerly successful
constituent quark models lose predictability in the vicinity of
two-meson thresholds. The emergence of non-negligible effects due to
meson loops requires an explicit evaluation of the interplay between
$Q\bar q$ and $(Q\bar q)(q\bar q)$ Fock components. In contrast to the
$c\bar s$ sector, there is no experimental evidence of $J^P=0^+,1^+$
bottom-strange states yet. Motivated by recent lattice studies, in
this work the heavy-quark partners of the $D_{s0}^\ast(2317)$ and
$D_{s1}(2460)$ states are analyzed within a heavy meson chiral unitary
scheme. As a novelty, the coupling between the constituent quark model
P-wave $\bar B_s$ scalar and axial mesons and the $\bar B^{(\ast)}K$ channels is
incorporated employing an effective interaction, consistent with heavy
quark spin symmetry, constrained by the lattice energy
levels.

\end{abstract}

\maketitle
\section{Introduction}

The low-lying positive parity charmed-strange spectrum moved into
spotlight in 2003 after the experimental observation of the
$D_{s0}^{\ast}(2317)$ and $D_{s1}(2460)$ states by the \textsc{BaBar}~\cite{Aubert:2003fg}
and \textsc{Cleo}
collaborations~\cite{Besson:2003cp} in  the $D_s^{(\ast)+}\pi^0$
invariant mass spectrum with $J^P=0^+$ and $1^+$ quantum numbers,
respectively.  The interest in such states arose from their low masses
and widths, contrary to constituent quark
model (CQM)~\cite{Godfrey:1986wj, Zeng:1994vj, Ebert:1997nk,
  DiPierro:2001dwf} and lattice QCD (LQCD)~\cite{Boyle:1997rk,
  Hein:2000qu, Bali:2003jv, Dougall:2003hv} expectations.  Consequently, the nature
    of those resonances became a matter of intense research, mainly
    being interpreted as naive $c\bar s$~\cite{Fayyazuddin:2003aa, Sadzikowski:2003jy,
      Lakhina:2006fy} or meson-meson and four-quark
    states~\cite{Barnes:2003dj,Lipkin:2003zk,Szczepaniak:2003vy,Browder:2003fk,Bicudo:2004dx,Gamermann:2006nm,MartinezTorres:2011pr,Albaladejo:2016hae}. An indirect hint for the non-perturbative
nature of the $D K$ chiral amplitudes near threshold, and the
possible existence of a bound state, was
obtained in Ref.~\cite{Flynn:2007ki} from an Omn\`es analysis of the LQCD data
for the scalar form factor in the $D\to \bar K \ell  \bar\nu_\ell$
semileptonic decay.
More recently, lattice QCD
simulations~\cite{Mohler:2013rwa,Lang:2014yfa,Torres:2014vna} and the quark model 
calculations~\cite{vanBeveren:2003jv,Ortega:2016mms} emphasized the
importance of including the $D^{(\ast)}K$ dynamics when describing the
P-wave $c\bar s$ mesons.  These meson-meson channels produce large
mass shifts which improve the  description of the experimental masses.

The combination of a heavy quark and a light antiquark in the $D_s$ or
$\bar B_s$ spectra is a great advantage when it comes to describe the
system. In such mesons, heavy quark spin symmetry
(HQSS)~\cite{Isgur:1989vq, Isgur:1989ed, Georgi:1990um, Isgur:1991wq} (see also the book  \cite{Manohar:2000dt}) is in good approximation fulfilled by QCD,
being thus the parity and the total angular momentum of the light
antiquark $j_{\bar q}$ approximately conserved. The decoupling of the spin of
the heavy quark ($s_Q$) and $j_{\bar q}$ generates the arrangement of states
in doublets labelled by their parity\footnote{Note that the parity of
  the light degrees of freedom coincides with that of the meson.} and $j_{\bar q}$ values, so members
within a doublet are governed by the same dynamics and become
degenerate in mass, up to $\Lambda_\text{QCD}/m_Q$ corrections, with $m_Q$
the heavy quark mass and $\Lambda_\text{QCD}$, a typical scale
accounting for the dynamics of the light degrees of freedom.  The
$D_{s0}^\ast(2317)$ and $D_{s1}(2460)$ are members of a
positive-parity $j_{\bar q}=\tfrac{1}{2}$ doublet, which will 
strongly couple to S-wave $D^{(\ast)}K$ pairs, being the dynamics of
these latter meson pairs  in turn governed by chiral symmetry. Within
a CQM scheme, these resonances will correspond to
P-wave states, where the spin and angular momentum of the light antiquark
are coupled to a total $j_{\bar q}=\tfrac{1}{2}$. There will be another HQSS
doublet with $j_{\bar  q}^{P}=\tfrac{3}{2}^{+}$.

Besides, in the $m_Q\to\infty$ limit  the dynamics of  systems containing a
single heavy quark  becomes also independent of the flavor of the
heavy quark~\cite{Manohar:2000dt}.\footnote{Note that, there appears also an
  approximate SU(3) flavor symmetry in the light sector.} Hence the bottom-strange
sector is expected to present a pattern similar to that of the $c\bar
s$ one, and in particular there should exist heavy-flavor partners of
the $D_{s0}^\ast(2317)$ and $D_{s1}(2460)$  resonances.  Furthermore,
since the $b$ quark is heavier than the $c$ quark, the 
$\mathcal{O}(m_Q^{-1})$ corrections are expected to be smaller, and
thus the HQSS relations should be more accurate. In 
these circumstances, the $b\bar s$ $j_{\bar q}^P=1/2^{+}$ doublet is a
perfect scenario to discuss the interplay between CQM 
states and  meson-meson channels with thresholds located close to the former. This study is relevant to unveil
the nature of the $D_{s0}^\ast(2317)$ and $D_{s1}(2460)$, where such
interplay turns out to be essential to understand the dynamics of
these states. Hence, we expect also  a strong influence  of the
close continuum two-meson channels on the properties of the bottom
partners of these even-parity resonances.

Unfortunately, unlike the $c\bar s$ spectrum, the lowest-lying positive parity
$b\bar s$ states have not been fully discovered. While experimental searches
have successfully observed the $j_{\bar q}^{P}=\tfrac{3}{2}^{+}$
doublet $\bar B_{s1}(5830)$ and
$\bar B_{s2}^\ast(5840)$~\cite{Aaltonen:2007ah,Abazov:2007af}, the
lower mass $j_{\bar q}^{P}=\tfrac{1}{2}^{+}$ doublet states still
wait to be observed. Note that the dynamics of the $j_{\bar
  q}^{P}=\tfrac{3}{2}^{+}$ doublet is not governed by chiral symmetry,
since the $\bar B^{(\ast)}K^\ast$ channel, involving the light vector
meson $K^\ast$, should be taken into account.
 
Under this lack of experimental data, many theoretical predictions
have been produced within a wide variety of techniques (quark
models~\cite{DiPierro:2001dwf,Ebert:2009ua,Sun:2014wea}, effective
field theory approaches
(EFTs)~\cite{Kolomeitsev:2003ac,Bardeen:2003kt,Guo:2006fu,Guo:2006rp,Cleven:2010aw,Colangelo:2012xi,Altenbuchinger:2013vwa,Albaladejo:2016lbb},
and LQCD~\cite{Gregory:2010gm, Lang:2015hza}). A special attention
deserves the recent LQCD study of the even-parity isoscalar $b\bar s$
energy-levels carried out in Ref.~\cite{Lang:2015hza}. There, clear
signatures for the $\bar B_{s1}(5830)$ and $\bar B_{s2}^\ast(5840)$
are found above the $\bar B^{(\ast)}K$ thresholds. Below these
thresholds, two  QCD bound states are identified using a
combination of quark-antiquark and $\bar B^{(\ast)}K$ interpolating
fields, and assuming that the mixing with $\bar B_s^{(\ast)}\eta$ and
the isospin-violating decays $\bar B_s^{(\ast)}\pi$ are negligible. A
$J^P=0^+$ bound state, with mass $5.711\pm 0.023$ GeV (adding errors
in quadrature) is predicted and with some further assumptions, it is
also found a $1^+$ state with a mass of $5.750 \pm 0.025$
GeV~\cite{Lang:2015hza}.
 
In this work, we will pay attention to the $0^+$ and $1^+$ isoscalar
bottom-strange sector. We will use a heavy meson chiral unitary scheme
to describe the isoscalar S-wave elastic $\bar B^{(\ast)}K\to \bar B^{(\ast)}K$\, $T-$matrix. The scattering
amplitudes will be
obtained  by solving a renormalized Bethe--Salpeter equation (BSE)
with an interaction kernel
determined from leading order (LO) heavy meson chiral perturbation theory
(HM$\chi$PT)~\cite{Grinstein:1992qt,Wise:1992hn,Burdman:1992gh,Yan:1992gz}. We 
will couple the two-meson channels with the CQM P-wave $\bar B_s$ scalar and axial mesons  using an effective
interaction consistent with HQSS. To that end, we will follow the
scheme detailed in Ref.~\cite{Cincioglu:2016fkm}, where the $(D\bar
D^\ast+h.c.)$ two-meson channel was coupled to the $\chi_{c1}(2P)$ charmonium
state, and the consequences for the $X(3872)$ and its spin-flavor
partners, were examined.  Finally, we will use the LQCD energy levels
reported in Ref.~\cite{Lang:2015hza} to constrain the undetermined
low-energy-constants (LECs) of the present approach. As a final
outcome,  we will present 
robust predictions for the lowest-lying $b\bar s$ $J^P=0^+$ and $1^+$
states, that can serve as an important guidance for experimental
searches and to shed light into the situation in the analog charm
sector. 

\section{\boldmath Isoscalar $\bar B^{(\ast)}K\to \bar B^{(\ast)}K$ scattering}

\subsection{HQSS fields}

To study the $\bar B^{(\ast)}K$ interactions, we first introduce the matrix
field $H_a^{(Q)}$, 
\begin{eqnarray}\label{eq:HQSSfields1}
 H_a^{(Q)}&=&\frac{1+\slashed{v}}{2}\left(P^{\ast(Q)}_{a\,\mu}\gamma^\mu-P^{(Q)}_a\gamma_5\right),
\end{eqnarray}
which combines the isospin doublet and singlet of pseudoscalar
heavy-mesons $P_a^{(Q)}=\left(Q\bar u,Q\bar d, Q\bar s\right)$ fields
and their vector HQSS partners $P_a^{\ast(Q)}$. We use the isospin
convention $\bar u = \vert 1/2,-1/2\rangle$ and $\bar d = -\vert
1/2,+1/2\rangle$, which induces a minus sign between the $D^+$ (and
$\bar B^0$) particle and isospin states.  Besides $v$ is the
four-velocity of the mesons, and the vector field satisfies $v\cdot
P_a^{\ast(Q)}=0$. Note that in our normalization the heavy-light meson
field, $H^{(Q)}$, has dimensions of $E^{3/2}$ (see
Ref.~\cite{Manohar:2000dt} for details). This is because we use a
non-relativistic normalization for the heavy mesons, which differs
from the traditional relativistic one by a factor $\sqrt{M_H}$.

On the other hand, within the HQSS formalism the even parity
CQM bare $Q\bar q $ states, associated to the $j_{\bar
  q}^P=\tfrac{1}{2}^+$ HQSS doublet, are described by the matrix field
$J_a^{(Q)}$~\cite{Falk:1991nq},
\begin{eqnarray}\label{eq:HQSSfields2}
 J_a^{(Q)}&=&\frac{1+\slashed{v}}{2}\left(Y^{\ast(Q)}_{a\,\mu}\gamma_5\gamma^\mu+Y^{(Q)}_a\right), 
\end{eqnarray}
with $v^\mu Y^{\ast(Q)}_{a\,\mu}=0$. The $Y_a$ and $Y_a^{\ast}$ fields
respectively annihilate the $0^+$ and $1^+$ meson states belonging to the
$\tfrac{1}{2}^+$  doublet. 

Under a parity transformation we have
\begin{equation}
H^{(Q)}(x^0,\vec{x}) \to \gamma^0  H^{(Q)}(x^0,-\vec{x})\gamma^0,
\qquad v^\mu \to v_\mu. \label{eq:parity}
\end{equation}
The field $H_a^{(Q)}$ transforms as a $(2,\bar 3)$ under the heavy
spin $\otimes $ SU(3)$_V$ symmetry~\cite{Grinstein:1992qt},
this is to say:
\begin{equation}
H_a^{(Q)} \to S_Q \left( H^{(Q)} U ^\dagger\right)_a. \label{eq:hqsf}
\end{equation}
The hermitian conjugate field is defined by:
\begin{equation}
\bar H^{(Q)a} =\gamma^0 [H_a^{(Q)}]^\dagger \gamma^0, \label{eq:barstates}
\end{equation}
and transforms as~\cite{Grinstein:1992qt}:
\begin{equation}
\bar H^{(Q)a} \to  \left( U \bar H^{(Q)} \right)^a S^\dagger_Q. \label{eq:barhqsf}
\end{equation}
The matrix field $J_a^{(Q)}$ satisfies transformation 
relations identical to those in
Eqs.~(\ref{eq:parity})--(\ref{eq:barhqsf}). 

\subsection{Interactions}

In S-wave, the spin-parity quantum numbers of the $\bar B K$ 
$\left(\bar
B^{\ast}K\right)$ meson pair are $0^+ (1^+)$, while the light degrees of
freedom are coupled to
spin-parity $1/2^+$. As in Ref.~\cite{Lang:2015hza}, we will neglect the $\bar
B_s^{(\ast)}\eta$ channels, and thus the (elastic)  isoscalar $\bar B^{(\ast)}K \to \bar B^{(\ast)}K$ interaction potential $V(s)$  consists of a chiral contact term
[$V_\text{c}(s)$] plus the exchange [$V_\text{ex}(s)$] of a bare
$b\bar{s}$ state,
\begin{equation}\label{eq:fullpot}
V(s) = V_\text{c}(s) + V_\text{ex}(s)
\end{equation}
with $s$, the center of mass (c.m.) energy squared. At LO in the chiral
counting, $V_\text{c}(s)$ is given by the Weinberg-Tomozawa Lagrangian
that
reads~\cite{Grinstein:1992qt,Wise:1992hn,Burdman:1992gh,Yan:1992gz}
(omitting from now on the $(Q)$ superscript),
\begin{equation} \label{eq:Otransition1}
 \mathcal{L}=\frac{i}{2} {\rm Tr}\left(\bar H^a H_b
 v^\mu\left[\xi^\dagger\partial_\mu\xi+\xi\partial_\mu\xi^\dagger\right]_a^b\right),
\end{equation}
where the $\xi$ field contains the pseudoscalar light mesons, and can
be written as $\xi=\exp(\tfrac{iM}{\sqrt{2}f})$, with $f\sim 93$ MeV,
and the $M$ matrix defined in Ref.~\cite{Wise:1992hn}. The Lagrangian
density in Eq.~(\ref{eq:Otransition1}) is invariant
under ${\rm SU(3)}_L\times {\rm SU(3)}_R$ chiral symmetry, Lorentz
transformations, HQSS and parity. From the previous
Lagrangian, the isoscalar contact term contribution $V_\text{c}(s)$ can be
easily derived, and the result after projecting into S-wave
reads:\footnote{For $J=1$ there appears the product of the polarization
  vectors of the initial and final $\bar B^\ast$ mesons, which is
  approximated by $-1$, after neglecting corrections suppressed by
  the heavy meson mass.}
\begin{equation}\label{eq:contact}
V_\text{c}(s) = \frac{\displaystyle -3s + (m_{\bar B^{(\ast)}}^2-m_K^2)^2/s + 2(m_{\bar B^{(\ast)}}^2+m_K^2)}{4f^2}~.
\end{equation}
Neglecting the $\bar B^\ast-\bar B$ mass difference, the interactions
in the  $J=0$ and 1 sectors are identical, as expected from  HQSS
because they correspond to the same configuration ($1/2^+$) of the light
degrees of freedom.

The exchange term in Eq.~\eqref{eq:fullpot} is determined by the
coupling between the $\bar B^{(\ast)}K$ meson pairs and the P-wave bare quark
model states described by the matrix field $J_a^{(Q)}$ introduced in
Eq.~\eqref{eq:HQSSfields2}. At LO in the heavy quark expansion, there
exists only one term invariant under Lorentz, parity, chiral and heavy
quark spin transformations,
\begin{equation} \label{eq:Otransition2}
 \mathcal{L}=\frac{ic}{2} {\rm Tr}\left(\bar H^a J_b
 \gamma^\mu\gamma_5\left[\xi^\dagger\partial_\mu\xi-\xi\partial_\mu\xi^\dagger\right]_a^b\right)+h.c.~,
\end{equation}
where $c$ is a dimensionless undetermined LEC that controls the
strength of the vertex. This LEC, though it depends on the orbital
angular momentum and radial quantum numbers of the CQM state, is in principle independent of the spin of the
quark-model heavy-light meson, and of both the heavy-quark flavor and
the light SU(3) flavor structure of the vertex. Thus, up to
$\Lambda_{\rm QCD}/m_Q$ corrections, it can be used both for $J=0$ and
$J=1$ in the charm and bottom sectors. Moreover, the same LEC governs the interplay
between two-meson and quark model degrees of freedom in all isospin
and strangeness channels. Paying attention to the isoscalar
bottom-strange sector, we find a $b\bar{s} \to \bar B^{(\ast)}K$
coupling in S-wave,
\begin{equation}\label{eq:Vcoupling}
 V_{b\bar s}(s) = \frac{ic}{f}\sqrt{\frac{m_{\bar B^{(\ast)}} \mbare}{s}} \left(s-m_{\bar B^{(\ast)}}^{2}+m_K^2\right),
\end{equation}
where $\mbare$ is the  mass of the $b\bar s$ meson without the effect
of the $\bar B^{(\ast)}K$ meson loops. This mass is the same, up to small
HQSS breaking corrections, for both $J=0$ and $J=1$ sectors, and it can be in
principle obtained from CQMs.  In what follows, we
will denote it as the \emph{bare mass} of the state.\footnote{Owing to
  SU(3) light flavor-symmetry, the bare mass would present also a soft
  pattern of isospin and strangeness corrections.} Note that, here, by
  bare mass, we mean the mass of the CQM states when the LEC
  $c$ is set to zero, and thus it is not a physical observable. In the
  sector studied in this work, the coupling 
 to the $\bar B^{(*)} K$ meson pairs renormalizes this
  bare mass, as we will discuss below. Since, in the effective theory,
  the ultraviolet (UV) regulator is finite, the difference between the bare and the
  physical resonance masses is a finite renormalization. This shift
  depends on the UV regulator since the bare mass itself depends on
  the renormalization scheme.  The value of the bare mass, which is
  thus a free parameter, can either be indirectly fitted to
  experimental observations, or obtained from schemes that ignore the
  coupling to the mesons, such as some
  CQMs. In this latter case, the issue certainly
  would be to set the UV regulator to match the quark model and the
  EFT approaches. We will come back to this point later.

The Lagrangian density in Eq.~\eqref{eq:Otransition2} allows to compute
the $V_\text{ex}(s)$ term contribution to the full potential,
Eq.~\eqref{eq:fullpot}, that accounts for  $\bar B^{(\ast)}K$ scattering
via the exchange of intermediate even-parity bottom-strange mesons~\cite{Cincioglu:2016fkm}:
\begin{equation}
 \label{eq:exchange}
 V_\text{ex}(s)=\frac{V_{b\bar s}(s)V_{b\bar s}^\dagger(s)}{s-\mbare^2}~.
\end{equation}
The HQSS consistent potential $V(s)$ detailed above is used to obtain
the ${\bar B^{(\ast)}}K$ elastic unitary amplitude, $T(s)$. This is
done by 
solving the BSE within the so-called on-shell
approximation~\cite{Nieves:1999bx}. We use~\cite{Flynn:2007ki}:
\begin{equation}\label{eq:InvTMatrix}
 T^{-1}(s)=V^{-1}(s)-G(s)~,
\end{equation}
where $G(s)$ is the two-meson loop integral, regularized with a Gaussian cut-off,
\begin{eqnarray}
 G(s) &=& \int \frac{d^3q}{(2\pi)^3}
 \frac{\Omega(\vec{q}\,)
 e^{-2(\vec{q}^{\,2}-\vec{k}^{\,2})/\Lambda^2}}{s-(\omega(\vec{q}\,)+\omega'(\vec{q}\,))^2+i\epsilon}
 \\
&=& -i\frac{|\vec{k}\,|}{8\pi\sqrt{s}}\Theta\left[s-(m_{\bar B^{(\ast)}}+m_K)^2\right]\nonumber\\
&+& {\cal P}\left( \int \frac{d^3q}{(2\pi)^3}
 \frac{\Omega(\vec{q}\,)
 e^{-2(\vec{q}^{\,2}-\vec{k}^{\,2})/\Lambda^2}}{s-(\omega(\vec{q}\,)+\omega'(\vec{q}\,))^2}\right)~.\label{eq:Gloop1}
\end{eqnarray} 
Above, ${\cal P}(\cdots)$ stands for the principal value of the
integral and 
\begin{equation}
\Omega(\vec{q}\,)=
\frac{\omega(\vec{q}\,)+\omega'(\vec{q}\,)}{2\omega(\vec{q}\,)\omega'(\vec{q}\,)}~,
\end{equation}
with $\omega(\vec{q}\,)$ and $\omega'(\vec{q}\,)$  the energies of
the $\bar B^{(\ast)}$ and $K$ mesons, respectively. Finally,
$\vec{k}^{\,2}$ is  the square of the c.m. on-shell momentum,
\begin{equation}
\vec{k}^{\,2} = \frac{\left(s-s_+\right)\left(s-s_-\right)}{4s}, \quad
s_{\pm}=(m_{\bar B^{(\ast)}}\pm  m_K)^2~. \label{eq:defk2}
\end{equation}
\subsection{Bound, resonant states, couplings and the compositeness
  condition for bound states }

The dynamically-generated meson states appear as poles of the
scattering amplitudes on the complex $s-$plane. The poles of
the scattering amplitude on the first Riemann sheet (FRS) that appear
on the real axis below threshold, $s_+$, are
interpreted as bound states. The poles that are found on the second
Riemann sheet (SRS) below the real axis and above threshold are
identified with resonances. The SRS is simply obtained by analytical
continuation of the amplitude in the physical FRS
across the unitarity cut,
\begin{equation}
G_{\rm SRS}(s) =  G_{\rm FRS}(s)+ i\frac{k(s)}{4\pi\sqrt{s}},\quad
s\in \comp~,
\end{equation}
where
\begin{equation}
\frac{k(s)}{\sqrt{s}} =\frac{\left[\left(s-s_+\right)\left(s-s_-\right)\right]^\frac12}{2s}~.
\end{equation}
Note that the cuts for $k(s)/\sqrt{s}$  go along the
real axis for $-\infty < s < s_- $ and $ s_+ < s < \infty $. The
function is chosen to be real and positive on the upper lip of the
second cut, $ s_+ < s < \infty $, and hence it satisfies:
\begin{eqnarray}
0<\left. \frac{k(s)}{\sqrt{s}}\right|_{(s+{\rm i}
\epsilon)} = -\left. \frac{k(s)}{\sqrt{s}}\right|_{(s-{\rm i}
\epsilon)}, s_+< s  \in \reales~.
\end{eqnarray}
The mass and the width of the state can be found from the position of
the pole on the complex energy plane. Close to the pole, the
scattering amplitude behaves as
\begin{eqnarray}\label{eq:Tapprox}
T&\sim&\frac{g^2} {s-s_R}~.
\end{eqnarray}
The mass $M_R$ and width $\Gamma_R$ of the state result 
from $\sqrt{s_R} = M_R-i\ \Gamma_R/2$, while $g$ (complex in general) is the
coupling of the state to the ${\bar B^{(\ast)}}K$ channel.

Information on the compositeness of the bound states can be
obtained from the derivative of the meson loop function and the residue at the
pole position. Indeed, it can be
shown~\cite{Gamermann:2009uq,Aceti:2014ala}, inspired by 
  the Weinberg compositeness condition~\cite{Weinberg:1962hj, Weinberg:1965zz, Baru:2003qq},  that 
the probability of finding the ${\bar B^{(\ast)}}K$ molecular
component in the bound state wave function is given by
\begin{equation}\label{eq:Molprob}
 P_{{\bar B^{(\ast)}}K}=-g^2\left.\frac{\partial G}{\partial s}\right|_{s=M_b^2}~,
\end{equation}
where $M_b$ is the mass of the bound state and $g$ the coupling of the
state to  the ${\bar B^{(\ast)}}K$ meson pair. The above probability
deviates from one because of the energy dependence of the potential 
[Eq.~\eqref{eq:exchange}]~\cite{Aceti:2014ala,Garcia-Recio:2015jsa},
which is enhanced by the  exchange of intermediate quark model (bare)
bottom-strange mesons~\cite{Cincioglu:2016fkm}. We do not extend this
discussion to resonances, restricting it to the
case of bound states. For poles located in the complex plane 
the  strict probabilistic interpretation is lost, since the definition
in Eq.~\eqref{eq:Molprob} gives rise to complex numbers (see for
instance the discussion in Ref.~\cite{Cincioglu:2016fkm}).

\subsection{Finite volume}

To compare with LQCD simulations, we consider the
$T$-matrix [Eq.~\eqref{eq:InvTMatrix}] in a finite box of size
$L$. The boundaries of the box impose a quantization condition for the
momentum, $\vec{q} = \frac{2\pi}{L}\vec{n}$, with $\vec{n}\in
\mathbb{Z}^3$. The loop function $G(s)$ is thus replaced with its
finite volume counterpart, $\widetilde{G}(s,L)$~\cite{Doring:2011vk,Albaladejo:2013aka},
\begin{equation}\label{eq:Gfv}
\widetilde{G}(s,L) = \frac{1}{L^3}\sum_{\vec{n}\in \mathbb{Z}^3}
\frac{\Omega(\vec{q}\,) e^{-2(\vec{q}^{\,2}-\vec{k}^{\,2})/\Lambda^2}}{s-(\omega(\vec{q}\,)+\omega'(\vec{q}\,))^2}~.
\end{equation}
Up to the order we are considering in this work, the potential
does not receive  finite volume corrections, and hence the finite volume
$T$-matrix, denoted as $\widetilde{T}(s,L)$, reads:
\begin{equation}
\widetilde{T}^{-1}(s,L) = V^{-1}(s) - \widetilde{G}(s,L)~.
\end{equation}
The energy levels  obtained in LQCD simulations can be computed within
our approach from the poles of $\widetilde{T}(s,L)$. 

To better describe the
energy levels reported in the LQCD simulation carried out in 
Ref.~\cite{Lang:2015hza}, we  use
the masses and the energy-momentum dispersion relations
given in that work. In particular, we will employ a modified energy-momentum
dispersion relation for the $\bar B^{(\ast)}$ mesons,
\begin{equation}
\omega(\vec{q}\,) \to \omega^\text{lat}(\vec{q}\,) = m_1 +
\frac{\vec{q}^{\,2}}{2m_2} -
\frac{(\vec{q}^{\,2})^2}{8m_4^3}~, \label{eq:lqcd-disp} 
\end{equation}
where the parameters appearing in the above equation can be found in
Table 1 of Ref.~\cite{Lang:2015hza}.  The lattice size and spacing in
that simulation are $32^3\times 64$ and $a=0.0907\pm 0.0013$ fm,
respectively, while the pion mass is $156 \pm 7$ MeV. For the kaon,
the ordinary relativistic dispersion relation is used with an
unphysical mass of $m_Ka= 0.2317\pm 0.0006$ ($m_K=504\pm 7$ MeV)\cite{Lang:2014yfa}. To compute the potentials
(chiral+exchange) at finite volume, we set the $\bar B^{(\ast)}$ mass
to $m_1$, introduced in the modified dispersion relation of
Eq.~\eqref{eq:lqcd-disp}. Finally $\vec{k}^{\,2}$, that appears in the
Gaussian regulator needed to render $\widetilde{G}(s,L)$ finite is obtained from
Eq.~\eqref{eq:defk2} using the lattice masses.

\section{Results}
\begin{table*}
\caption{\label{tab:Results} Parameters of the model fitted to the
  energy levels of Ref.~\cite{Lang:2015hza}, together with masses and
  properties of the low-lying $j_{\bar q}^P=\tfrac{1}{2}^+$ $\bar B_s$
  meson doublets deduced from these parameters in the infinite volume
  case. In this latter case, physical $\bar B^{(\ast)}$ and $K$ masses
  have been used, and we have searched for poles in the FRS (bound
  states) and SRS (resonances) of the isoscalar S-wave $\bar
  B^{(\ast)} K$ amplitudes.  Besides, we find $a^{\rm th}=0.0909 \pm
  0.0013$ fm and $0.0910 \pm 0.0013$ fm for sets (a) and (b),
respectively, which compare rather well with the lattice spacing
  ($a=0.0907\pm 0.0013$ fm) quoted in Ref.~\cite{Lang:2015hza}. The
  isoscalar $0^+$ and $1^+$ scattering lengths ($a_0$) and the
  isoscalar S-wave $\bar B^{(\ast)} K\to \bar B^{(\ast)} K $
  amplitudes at threshold are related by $T(s_+)= -8\pi a_0
  \sqrt{s_+}$, with $s_+=(m_{\bar B^{(\ast)}}+m_K)^2$. The ${\bar
    B^{(\ast)}}K$ molecular probability $P_{\bar B^{(\ast)}K}$ is
  calculated using Eq.~\eqref{eq:Molprob} and it is given only for
  bound states. The coupling $g$, defined in Eq.~\eqref{eq:Tapprox},
  is also calculated only for the bound state. Errors on the fitted
  parameters show 68\% confidence levels (CLs), which are obtained
  from distributions computed by performing a large number of best
  fits to Monte Carlo synthetic datasets. The synthetic datasets are
  generated from the original energy levels of
  Ref.~\cite{Lang:2015hza} and the lattice spacing assuming that the
  data points are Gaussian distributed. The 68\% CLs are obtained by
  discarding the higher and the lower 16\% tails of the appropriate
  distributions. These parameter distributions are used to estimate
  the uncertainties on the derived quantities in the infinite volume
  case. }
\begin{ruledtabular}
\def\arraystretch{1.3}
\begin{tabular}{cccccc|cccccc}
\multicolumn{6}{c|}{Parameters} & \multicolumn{6}{c}{Infinite volume predictions}\\
   set   & $J^P$ & $\mbare$ [MeV] & $c$ & $\Lambda$ [MeV]& $\chi^2/\text{d.o.f.}$   & $M_b$ [MeV] &  $P_{\bar B^{(\ast)}K}$ [\%] & $g$ [GeV] & $a_0$ [fm] & $M_R$ [MeV] & $\Gamma_R$ [MeV]  \\  
  \hline
      \multirow{2}{*}{(a)} & $0^+$ & $5851$  &
      \multirow{2}{*}{$0.74\pm 0.05$} & \multirow{2}{*}{$730\pm 40$} & \multirow{2}{*}{$1.5$} 
& $5711\pm 6$ & $51.8\pm 1.5$ & $31.8 \pm 0.9$ &
      $-0.90\pm 0.05$ & $6300\pm 100$ & $70_{-40}^{+30}$\\
   & $1^+$ & $5883$ &  &  &  & 
   $5752\pm 6$  & $49.7\pm 1.4$ & $32.3\pm 0.9$ & $-0.87 \pm 0.04$
        &$6300\pm 100$ & $80_{-50}^{+30}$ \\  
  \hline
 \multirow{2}{*}{(b)} & $0^+$ & $5801$ &\multirow{2}{*}{$0.75\pm
   0.04$} & \multirow{2}{*}{$650  \pm 30$}  & \multirow{2}{*}{$1.6$}& 
 $5707 \pm 6$  & $45.8 \pm 1.1$ & $32.3 \pm 0.8$
 &$-0.88\pm 0.05$ &$6220\pm 70$ &$80_{-40}^{+30}$ \\
  & $1^+$ &  $5858$ &  &  &   & 
  $5757\pm 6$  & $48.3\pm 1.3$ & $32.3\pm 0.8$ &
 $-0.92\pm 0.05$ & $6280\pm 70$ & $70_{-40}^{+30}$\\
\end{tabular}
\end{ruledtabular}
\end{table*}
\begin{figure*}[t]
\includegraphics[width=0.49\textwidth,keepaspectratio]{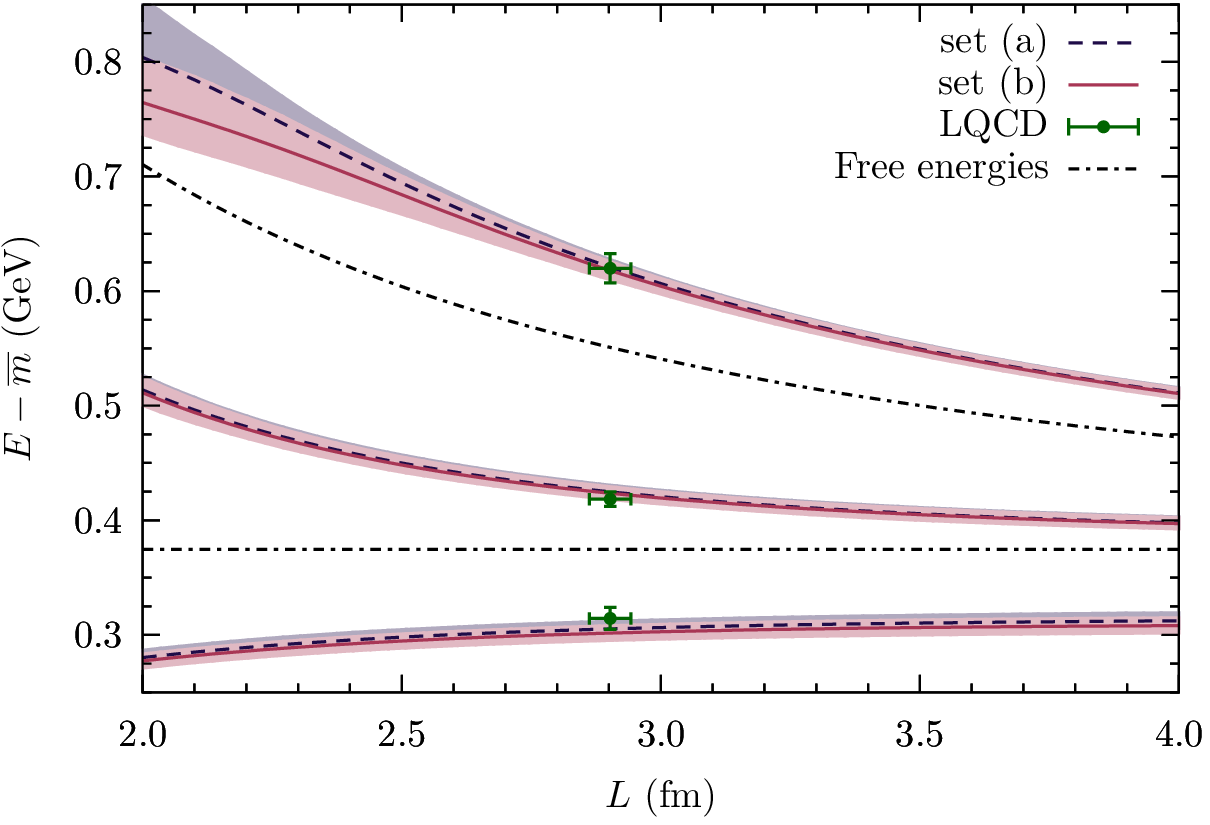}
\includegraphics[width=0.49\textwidth,keepaspectratio]{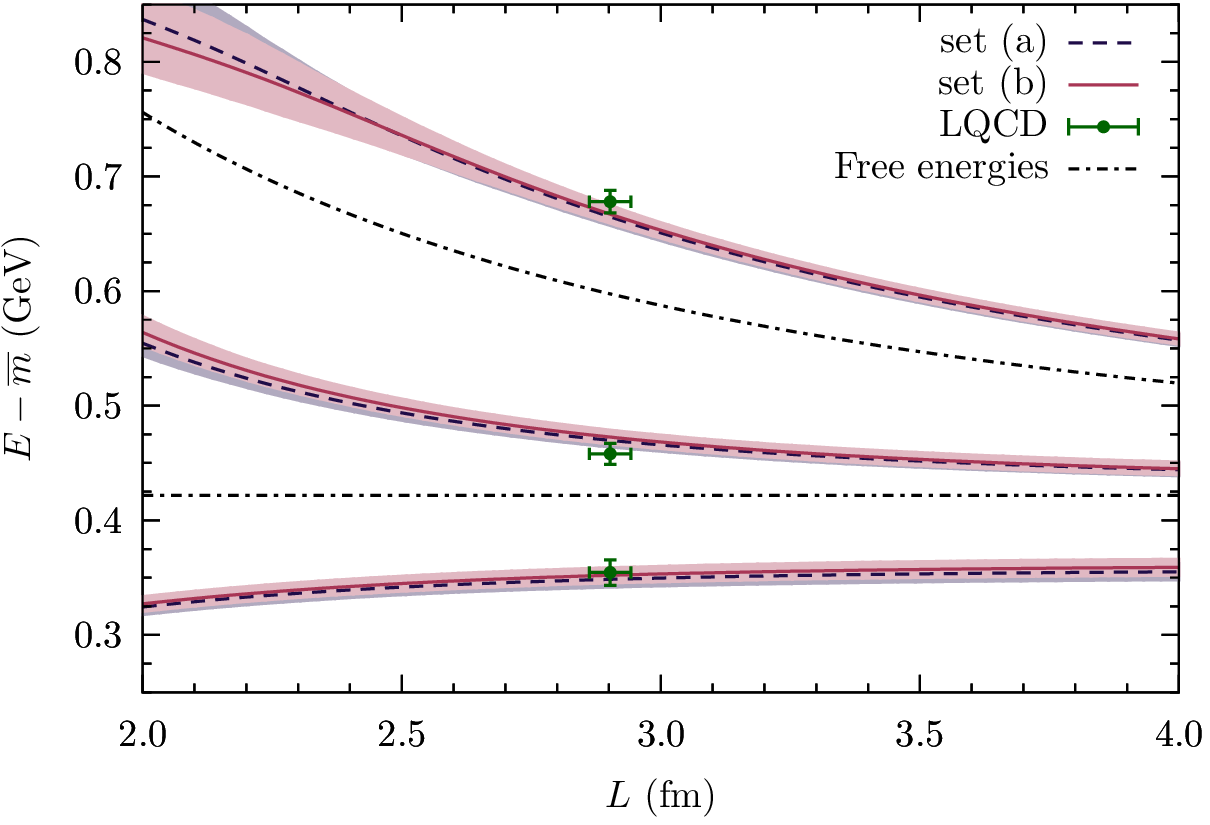}
\caption{\label{fig:wide} LQCD energy levels for $J^P=0^+$ (left) and
  $J^P=1^+$ (right), as a function of the box size $L$. We compare our
  predictions for the difference $E-\bar m$ with the results of
  Ref.~\cite{Lang:2015hza}, given also in Table 3 of that work. We
  have used $\bar m a=1.62897 (43)$, as in Ref.~\cite{Lang:2015hza}.  We
  have evaluated energy levels ($\widetilde{T}(s, L)$ presents poles
  at $\sqrt{s}=E$ for a given $L$, when the lattice masses are
  employed) using both sets of parameters compiled in
  Table~\ref{tab:Results}, which are shown by dashed blue and solid
  red lines for sets (a) and (b), respectively. Black lines stand for
  the first and second non-interacting $\bar B^{(\ast)}K$ levels,
  while the data points show the lattice levels reported in
  Ref.~\cite{Lang:2015hza}. The uncertainty bands in the predicted
  energy levels mark 68\% CLs obtained from a Monte Carlo simulation
  considering the parameter distributions of Table~\ref{tab:Results}.
}
\end{figure*}
We will fit our model to the levels 1,2,3 and 1,3,4 given in Table 3
of Ref.~\cite{Lang:2015hza} for the $0^+$ and $1^+$ sectors,
respectively.\footnote{Note that the $1^+$ level 2 is interpreted in 
\cite{Lang:2015hza}  to be the $j_{\bar q}^{P}=\tfrac{3}{2}^{+}$ state
with $J^P=1^+$, which does not couple to $B^* K$ in S-wave in the
heavy quark limit.} We will consider the energy-levels in lattice units. Hence, and to
properly take into account the uncertainty on this scale, we introduce
it as an additional best-fit parameter, $a^{\rm th}$, constrained by the central
value and error quoted above and taken from Ref.~\cite{Lang:2015hza}. 
Thus, we  minimize the following $\chi^2$,
\begin{equation}
\chi^2 = \sum_i \left( \frac{(Ea)^{\rm lat}_i-(Ea)^{\rm
    th}_i}{\Delta\left[(Ea)^{\rm lat}_i\right] }\right)^2 +
\left(\frac{a^{\rm lat}-a^{\rm th}}{\Delta\left[a^{\rm lat}\right]
}\right)^2~. \label{eq:defchi2}
\end{equation}
The sum runs over the six $0^+$ and $1^+$ energy-levels (given in
lattice units) specified above.  We could instead have fitted directly
to the energy levels in physical units, but in that case, the errors
on the levels inherited from the lattice spacing need to be treated as
correlated ones, since variations in the lattice spacing affect to all
the energy levels in the same manner. In addition to the use of
correlated errors, one would have also to consider that the lattice
meson masses, appearing in the theoretical $\widetilde{T}(s, L)$, will
change with $a$ as well because of their conversion into physical units. 
The $\chi^2$ introduced in
Eq.~\eqref{eq:defchi2} accounts for all these correlations. Indeed,
the latter  make the uncertainties on $(Ea)^{\rm lat}_i$ to become the most relevant ones. Note that
if these correlations induced by the lattice spacing are not taken
into account, one will end up with large and unrealistic
errors. On the other hand, one might treat $m_Ka$,
$(m_{1,2,4})a$ also as best-fit parameters, supplementing
appropriately the $\chi^2$. Results do not change significantly and
for simplicity we have fixed all these masses in lattice units to the
central values reported in Ref.~\cite{Lang:2015hza}. This little
sensitivity can be expected since the errors on $m_Ka$ and $m_1a$,
which determine the threshold and the chiral potential, are much
smaller than those on $a$. Indeed, the largest uncertainty in the
magnitude of these quantities is induced from the error on the lattice
spacing.

 Besides the lattice
spacing, the parameters of the model are the bare masses of the CQM $b\bar s$ $0^+$
and $1^+$ mesons, $\mbare$, the LEC $c$ that gives us the strength of
the coupling of the latter states with the two-meson $\bar
B^{(\ast)}K$ channels, and the UV Gaussian regulator $\Lambda$. We
expect to get reasonable estimates for the bare masses from the
predictions of
CQMs~\cite{DiPierro:2001dwf,Ebert:2009ua,Sun:2014wea}. These kind of
models find masses in the ranges $5800-5850$ MeV and $5840-5880$ MeV
for the $J^P=0^+$ and $J^P=1^+$ sectors, respectively. The $\bar B K$
and $\bar B^{\ast}K$ thresholds are located approximately at 5775 MeV
and 5820 MeV, respectively. Thus, in principle, we expect the quark
model states to be relatively close to, but above, the respective
$\bar B^{(\ast)}K$ thresholds, which would produce attractive $\bar
B^{(\ast)}K$ interactions for energies below the bare
masses~\cite{Cincioglu:2016fkm}.

We will explore the ranges of bare masses mentioned above, and we will
perform two independent fits using $\mbare$ values close to their
respective lower and upper limits. To maintain a consistent $0^+-1^+$
mass splitting, we will use the predictions of a widely used
non-relativistic
CQM~\cite{Vijande:2004he,Valcarce:2005em,Segovia:2013wma}.  This quark
model was already employed to study the low-lying P-wave
charmed-strange mesons~\cite{Ortega:2016mms}. In that reference, since
the $D^*_{s0}$ $1^3P_0$ ($n ^{2S+1}L_J$) bare state was found
significantly above the experimental level (2511 versus 2317.7 MeV),
an additional one-loop correction to the one-gluon exchange (OGE)
potential was introduced. This extra term was motivated from the
studies of Refs.~\cite{Gupta:1994mw,Lakhina:2006fy}, where a
spin-dependent term was added to the quark-antiquark potential
affecting only mesons in the case of unequal quark masses.  Such
correction is in general rather small, except for the $0^+$ sector,
where a large mass shift was found (around 128 MeV in the case of the
$D^*_{s0}$).  Hence, as commented above, we will consider two sets of
bare masses $\mbare$. In the set (a), we will use bare masses of 5851
MeV and 5883 MeV for the $0^+$ and $1^+$ states, respectively, as
deduced from the CQM of
Refs.~\cite{Vijande:2004he,Valcarce:2005em,Segovia:2013wma} without
including any correction to the OGE potential. For the second set,
(b), we will fix the $0^+$ and $1^+$ bare masses to 5801 MeV and 5858
MeV, as obtained when the latter CQM is supplemented with the OGE
one-loop terms discussed in
Ref.~\cite{Lakhina:2006fy,Gupta:1994mw,Ortega:2016pgg}.  Since the
LQCD simulation carried out in Ref.~\cite{Lang:2015hza} uses
non-physical meson masses, the CQM bare masses have been corrected
using the difference between the experimental and the prediction of
Ref.~\cite{Lang:2015hza} for the spin-average mass $\bar
m=\tfrac{1}{4}(m_{\bar B_s}+3m_{\bar B_s^\ast})$.

For each set of bare masses, the values of the other two parameters,
$c$ and $\Lambda$, are obtained from a simultaneous fit to the first
three $J^P=0^+$ and $1^+$ energy levels reported in the LQCD study of
Ref.~\cite{Lang:2015hza}.  In Fig.~\ref{fig:wide}, we present the predicted energy levels as a function of
the lattice size $L$, and the values of the fitted parameters are compiled in
Table~\ref{tab:Results}. As can be seen, we find an excellent description of the LQCD levels of
Ref.~\cite{Lang:2015hza} in both $J^P$ sectors, despite  the large
deviations from the free levels \cite{Albaladejo:2016jsg}. The LEC $c$ is rather insensitive to the
used bare masses, taking a value about 0.75 with an error of around
6\%. The central values of the UV regulator present however a mild
dependence, and we find $\Lambda=730\pm 40$ MeV for set (a), while for set (b)
the fitted value is $\Lambda=650\pm 30$ MeV. We should remind that the CQM bare masses depend on the renormalization scheme, in particular
on the UV regulator, or equivalently the UV regulator is expected to
depend on the bare masses. Nevertheless,  set (a) and (b) UV
regulators  turn out to be almost compatible within errors.

When the loop function is renormalized by a suitable subtraction,  instead of using a Gaussian regulator, the physical results showed in Table~\ref{tab:Results} and Fig.~\ref{fig:wide} do not appreciably change, besides
some variation of the renormalization-scheme-dependent low energy constant $c$. Thus, a similar good reproduction of the LQCD energy levels is achieved. Note that the finite volume loop function, in both renormalization schemes, is related to the  L\"uscher function~\cite{Luscher:1986pf,Luscher:1990ux}, as it is shown in Refs.~\cite{Doring:2011vk,Albaladejo:2013aka}. Hence, the continuous volume dependent curves in Fig.~\ref{fig:wide} are essentially the L\"uscher curves obtained from the phase shift by solving
\begin{equation}
\delta(q) + \phi(\hat q) = n\pi
\end{equation}
with $\hat q=q L/2\pi$ and $\phi(\hat q)$ determined by the L\"uscher function (see Eq.~(6.13) of Ref.~\cite{Luscher:1990ux}). 
 
Next, and once the parameters have been fixed, we  search for poles in the FRS
(bound states) and SRS (resonances) of the  isoscalar S-wave $\bar B^{(\ast)} K$
amplitudes for the infinite volume case. Pole positions are also
compiled in Table~\ref{tab:Results}, together with the $0^+$ and $1^+$
isoscalar
scattering lengths and the probabilities of the molecular $\bar
B^{(\ast)} K$ component in the bound states. For both sets of
parameters, and for both $J^P=0^+$ and $1^+$ sectors, we find  a bound state (FRS)
and a resonance (SRS).

In all cases, the mass of the bound state is rather independent of the
UV regulator, or equivalently of the bare quark model mass, and it is
located more than 100 MeV below the corresponding bare pole,
consequence of the strong attraction produced by the chiral
potential. This is a first hint of the importance of the meson loops
in the dynamics of the bound state, which can be also inferred from
its large ($\sim 50\%$) molecular component.  From the  results of
the Table~\ref{tab:Results}, we predict masses of $5709\pm 8$ MeV
and $5755\pm 8$ MeV for the $\bar B^\ast_{s0}$ and $\bar B_{s1}$
states, respectively.  These states are the heavy flavor partners of
the charmed-strange $D^\ast_{s0}(2317)$ and $D_{s1}(2460)$ resonances,
and are clear candidates for future experimental searches.
The masses obtained in this work are in excellent agreement with the estimations given in
Ref.~\cite{Lang:2015hza}, and mentioned in the
introduction.\footnote{Note that the uncertainties obtained here are smaller than
those quoted in Ref.~\cite{Lang:2015hza} because we go beyond the effective
range approximation and determine a
potential (see the discussion in Ref.~\cite{Albaladejo:2013aka}).} They are
also quite compatible within errors with other HM$\chi$PT predictions, where
the explicit coupling (LEC $c$) of the two-meson channel and the bare
quark model state is not considered\footnote{In these schemes, such
  effect is encoded either in the renormalization subtraction
  constants or in higher order LECs, appearing at next-to-leading
  order (NLO) in the HM$\chi$PT expansion. (Note that in the present work,
  we obtain reasonable UV cutoff values $\sim 700$ MeV, which do not
  hide large higher order contributions~\cite{Albaladejo:2016eps,Guo:2016nhb}.)
  Despite the bare quark model pole was located above, and relatively
  close to,  
  the $\bar B^{(*)}K$ threshold, we find the bound state significantly
  below $\mbare$. Hence, the bare pole induces a mild energy
  dependence in the vicinity of the physical bound states, which can
  be accounted for by means of local terms in the
  potential~\cite{Cincioglu:2016fkm}. The bare pole, however, should
  produce a relevant energy dependence in the amplitudes above
  threshold and close to its
  position. }~\cite{Altenbuchinger:2013vwa,Cleven:2010aw,Guo:2006fu,Guo:2006rp,Albaladejo:2016lbb}.
In all cases a similar binding energy around $60-70$ MeV is obtained,
which favors a molecular interpretation of such states, where 
one would expect a $(\bar B_{s1}-\bar B^*_{s0})$ mass splitting
similar to the  $(\bar B^*-\bar B)$ one. The latter is around 45
MeV, while in our calculation we find $(m_{\bar B_{s1}}-m_{\bar B^\ast_{s0}})
\sim 41 $ MeV for set (a) and $\sim 49$ MeV for set (b), around
$4\ \text{MeV}$ below and above the pseudoscalar-vector mass splitting,
respectively. This is a clear indication of the non-canonical quark
model nature of the $\bar B_{s1}$ and $\bar B^*_{s0}$ states. It is interesting, though, to note that
the molecular proportion in the $0^+$ state ($\sim 50\%)$ is below the
EFT estimations for the $D_{s0}^\ast(2317)$, predicted to be around
$70\%$ of
$D^{(\ast)}K$~\cite{Liu:2012zya,Torres:2014vna,Albaladejo:2016hae}.

In Ref.~\cite{Ortega:2016pgg} two-meson loops and CQM bare poles are
also coupled. For the latter, the values of the bare masses are the
same as those used here. The $\bar{B}^{(\ast)} K$ interactions are
derived from the same CQM used to compute the bare states, instead of
using HM$\chi$PT. The $^3 P_0$ model is employed to couple both types
of degrees of freedom, and the quark model wave functions provide
form-factors that regularize the meson loops. The $0^+$ and $1^+$
states reported in Ref.~\cite{Ortega:2016pgg} are around
$30-40\ \text{MeV}$ less bound than those found here and in the LQCD
study of Ref.~\cite{Lang:2015hza}. Presumably, this is because the
$\bar{B}^{(\ast)}K$ interactions derived in the CQM of
Ref.~\cite{Ortega:2016pgg} are weaker than those used here. Molecular
probabilities are reported in Ref.~\cite{Ortega:2016pgg} to be around
$30-40\%$, which are smaller than those found in the present approach.

Regarding the isoscalar scattering lengths, we predict (combining the
results of both sets)  $a_0=-0.89\pm 0.07$ fm for both $J^P=0^+$ and $1^+$ sectors,
which compares well with the results $a_0^{\bar B K}=-0.85 \pm 0.10$
fm and $a_0^{\bar B^\ast K}=-0.97 \pm 0.16$ fm, obtained in the
analysis carried out in the LQCD study of Ref.~\cite{Lang:2015hza}. In
the approach of Ref.~\cite{Ortega:2016pgg}, the $0^+$ and $1^+$
scattering lengths turn out to be $\sim -1.18\ \text{fm}$ and
$-1.35\ \text{fm}$, respectively, which are larger (in absolute value)
than those found here and in Ref.~\cite{Lang:2015hza}. This is
expected, since the bound states in Ref.~\cite{Ortega:2016pgg}
lie closer to the respective thresholds.

We now pay attention to the extra poles found in the SRS, located well
above ($\sim$ 400-500 MeV) their respective thresholds. From the very
beginning we should take these results with some caution, since most
likely they should be affected by sizable NLO and higher threshold-channels corrections. In other words,
they are not as theoretically robust as those concerning the
lowest-lying $\bar B_{s1}$ and $\bar B^*_{s0}$ states. These
resonances, likely, are originated from the bare $b\bar s$-quark-model
poles that are dressed by the $\bar B^{(\ast)}K$ meson loops.  In that
case, the bare pole has been highly renormalized, moving to
significant higher masses ($\sim 6.2-6.3$ GeV) and acquiring a significant
$\bar B^{(\ast)}K$ width ($\sim 70-80$ MeV). We should also bear in mind
that radial excitations of the CQM states or $\bar
B^{(\ast)}K^{\ast}$ two-meson loops, neither of them taken into account in this
study, might lie in this region of energies. Further theoretical and
experimental studies will be helpful in shedding light on the possible
existence and properties of these resonant states.

\section{Conclusions}

We have adopted a chiral unitary approach, based on leading-order
HM$\chi$PT $\bar B^{(\ast)} K$ interactions, and for the very first
time in this context, the two-meson channels have been coupled to the
corresponding CQM P-wave $\bar B_s$ scalar and
axial mesons 
using an effective interaction consistent with HQSS. We have examined
two different sets of masses for the bare quark model poles, and in
each case, successfully fitted the rest of parameters to the recently
reported LQCD  isoscalar $b\bar s$
 $0^+$ and $1^+$  energy-levels~\cite{Lang:2015hza}. Results turned out to be rather
independent of the  bare masses, showing that the 
changes can be easily re-absorbed by means of reasonable variations
of  the UV regulator.

We have focused on the scalar and axial $\bar B^*_{s0}$ and $\bar
B_{s1}$ states, which form a HQSS $j_{\bar q}^P=1/2^{+}$ meson
doublet, heavy-flavor partner of that in the charmed-strange sector
integrated by the $D^\ast_{s0}(2317)$ and $D_{s1}(2460)$ resonances.
We have searched for bound states (poles in the FRS) of the  isoscalar S-wave $\bar B^{(\ast)} K$
amplitudes and found masses of $5709\pm 8$ MeV ($0^+)$ and
$5755 \pm 8 $ MeV ($1^+$) for these states.  Therefore, the $\bar B^*_{s0}$ and $\bar
B_{s1}$ appear  well below the $\bar
B K$ and $\bar
B^\ast K$ thresholds, being in this way the lowest-lying mesons
with these quantum numbers and stable through strong interactions.
These states are clear candidates for experimental search
in the LHCb experiment, $B-$factories or future high-luminosity
proton-antiproton experiments.

 We have also predicted  the isoscalar
elastic S-wave $\bar B K$ and $\bar B^\ast K$ scattering lengths to be similar and
approximately equal to $-0.89\pm 0.07$ fm, in good agreement with
the findings of Ref.~\cite{Lang:2015hza}.

We have obtained sensible UV cutoff values $\sim 700$ MeV, which do
not hide large higher order contributions. In addition, and within the
renormalization scheme adopted in this work, we have determined the
dimensionless LEC $c$ that controls the strength of the coupling
between the $\bar B^{(\ast)}K$ meson pairs and the P-wave bare quark
model states. This LEC, though it depends on the orbital angular
momentum and radial quantum numbers of the CQM state, is in principle
independent of both the heavy-quark flavor and the light SU(3) flavor
structure of the vertex. Thus, up to $\Lambda_{\rm QCD}/m_Q$
corrections, it can for example be also used to address the 
interplay between meson-loops and CQM degrees of freedom in the case
of the $D^\ast_{s0}(2317)$ and $D_{s1}(2460)$ resonances.  Moreover,
the same LEC governs the interplay between two-meson and quark model
degrees of freedom in all isospin and strangeness channels. Next, we
have looked at the Weinberg compositeness condition.  Thanks to this
admixture between CQM and two-meson degrees of freedom, we could
realistically estimate the molecular component ($\bar B^{(\ast)} K$)
of the $\bar B^*_{s0}$ and $\bar B_{s1}$, which turned out to be of
the order of $50\%$. This is a clear indication of the non-canonical
quark model nature of these states.

Finally, we have further predicted the volume dependence of the isoscalar
$b\bar s$ $0^+$ and $1^+$ energy-levels, which could
be useful for future LQCD simulations.


\begin{acknowledgments}
M. A. acknowledges financial support from the
``Juan de la Cierva'' program (27-13-463B-731) from the
Spanish MINECO. P. F.-S. acknowledges financial support from the "Ayudas para contratos predoctorales para la formaci\'on de doctores" program (BES-2015-072049) from the Spanish MINECO and ESF. This work is supported by the Spanish MINECO and European FEDER funds under the contracts FIS2014-51948-C2-1-P, FIS2014-57026-REDT and SEV-2014-0398 and by Generalitat Valenciana under contract PROMETEOII/2014/0068.
\end{acknowledgments}

\bibliography{BsHQSS}
\end{document}